\documentclass[sigplan,twocolumn]{acmart}

\pdfoutput=1

\acmConference[Preprint]{Preprint}{November 2023}{arXiv}
\renewcommand\footnotetextcopyrightpermission[1]{}
\usepackage{microtype}
\usepackage{booktabs}
\usepackage{hyperref}
\usepackage{color,soul}
\usepackage{graphicx}
\usepackage{xcolor}
\usepackage{tikz}
\usepackage{xspace}
\usepackage{tabularx}
\usepackage{subfig}
\usepackage{enumitem}
\usepackage{mdwlist}
\usepackage{xr}
\usepackage{multirow}

\graphicspath{ {./figs/} } 

\newcommand*\circled[1]{\tikz[baseline=(char.base)]{
\node[shape=circle,draw,inner sep=0.5pt,scale=0.8] (char) {#1};}}

\begin{document}

\title{Technical Report \\
       Cascade: A Platform for Delay-Sensitive Edge Intelligence}

\author{Weijia Song}
\orcid{0000-0003-2108-4998}
\affiliation{
    \institution{Cornell University}
    \country{USA}
}
\email{wsong@cornell.edu}

\author{Thiago Garrett}
\orcid{0000-0002-7171-5463}
\affiliation{
    \institution{University of Oslo}
    \country{Norway}
}
\email{thiagoga@ifi.uio.no}

\author{Yuting Yang}
\affiliation{
    \institution{Cornell University}
    \country{USA}
}
\email{yy354@cornell.edu}

\author{Mingzhao Liu}
\affiliation{
    \institution{Cornell University}
    \country{USA}
}
\email{ml2579@cornell.edu}

\author{Edward Tremel}
\affiliation{
    \institution{Augusta University}
    \country{USA}
}
\email{etremel@augusta.edu}

\author{Lorenzo Rosa}
\affiliation{
    \institution{University of Bologna}
    \country{Italy}
}
\email{lorenzo.rosa@unibo.it}

\author{Andrea Merlina}
\affiliation{
    \institution{University of Oslo}
    \country{Norway}
}
\email{andremer@ifi.uio.no}

\author{Roman Vitenberg}
\affiliation{
    \institution{University of Oslo}
    \country{Norway}
}
\email{romanvi@ifi.uio.no}

\author{Ken Birman}
\affiliation{
    \institution{Cornell University}
    \country{USA}
}
\email{ken@cs.cornell.edu}

\renewcommand{\shortauthors}{Song et al.}

\begin{abstract}
Interactive intelligent computing applications are increasingly prevalent, creating a need for AI/ML platforms optimized to reduce per-event latency while maintaining high throughput and efficient resource management. Yet many intelligent applications run on AI/ML platforms that optimize for high throughput even at the cost of high tail-latency.  Cascade is a new AI/ML hosting platform intended to untangle this puzzle.  Innovations include a legacy-friendly storage layer that moves data with minimal copying and a ``fast path'' that collocates data and computation to maximize responsiveness.  Our evaluation shows that Cascade reduces latency by orders of magnitude with no loss of throughput.
\end{abstract}

\maketitle

\begin{figure*}
  \centering
  \subfloat {   \includegraphics[width=0.5\linewidth,]{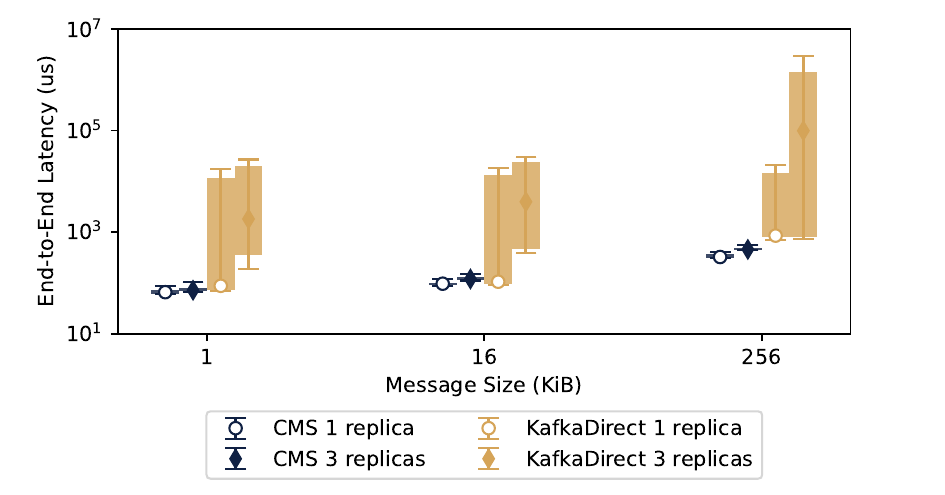}  }
  \subfloat{  \includegraphics[width=0.5\linewidth,]{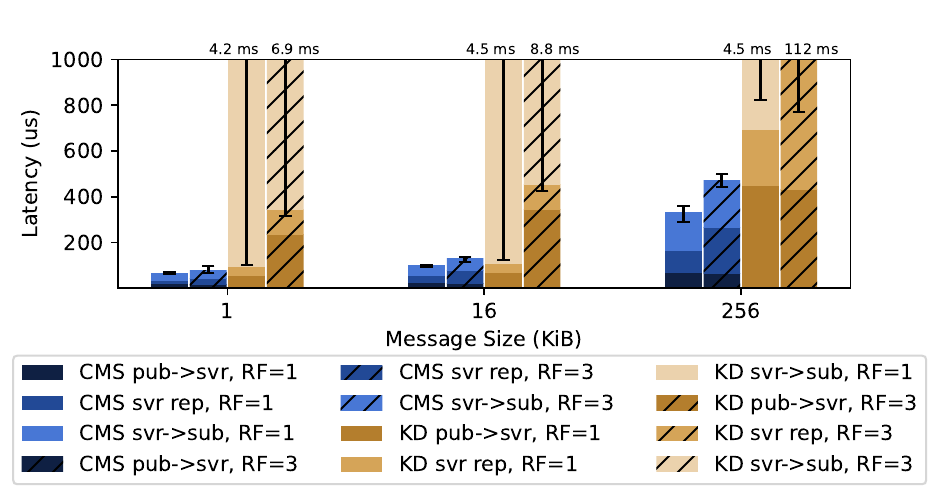}  }
\caption{Comparison of the Cascade messaging system (CMS) and Kafka Direct over 100Gbps RDMA.  For a two-stage no-op task, Cascade latency is less than 100{\em us}, whereas Kafka-Direct has huge tail latencies (left), and they arise at every layer (right).  The notation to describe the layers and full details of the experiment are presented in Sec.~\ref{sec:kafka-direct}}
\label{fig:kafka-direct}
\end{figure*}
\section{Introduction}
\label{sec:intro}

There is growing interest in intelligent edge applications that leverage low-latency edge networking for faster inference.  The end user gains snappier responses for time-sensitive tasks like interactive chat, medical diagnostics, and identifying defects in automated manufacturing settings. Meanwhile, the system itself can use intelligence too, for example to rapidly discard uninteresting images~\cite{aviation,locovolt,knight2021ge,hannon2021real}.
Microsoft predicts that all its end-user interactions will be AI-enhanced within a few years, a widely-shared goal.

Hardware advances are reducing compute latency in the MLs themselves. 
The challenge is to preserve these benefits and leverage recent advances without needing to reimplement existing high-value solutions.  For example, there has been a burst of work on improved ML runtimes and schedulers, often without requiring code changes~\cite{osdi2022yu,osdi2022han,osdi2023zhao,osdi2023chen}. Our effort is of a similar nature but focuses on the hosting
infrastructure: want to show that by improving hosting we can make it easier to build and deploy edge intelligence, and far more cost-effective to own such solutions.

Edge intelligence developers have a diversity of choices, but surprisingly many prioritize throughput over latency.  Indeed, our work started with a recognition that today's model servers have high-latency components on their critical paths, tracing to APIs like REST and gRPC: stage-to-stage handoffs that often add more delay than  ML computation.  But the issue is seen with other options too, such as interconnects based on Kafka pub/sub or stream processing tools like Spark Streaming, Google DataFlow and Apache Flink.  

Our ML hosting platform, Cascade, prioritizes low latency while treating high throughput and efficient use of hardware as important secondary objectives.  If available, inter-node communication leverages remote DMA networking (RDMA), but Cascade can also run on other fast network stacks like the data plane developer kit (DPDK). Cascade hosts data and computation within a single set of dual-capable servers, enabling collocation of compute and data to achieve ultra-low delay for time-sensitive tasks.  The data model centers on a distributed K/V store sharded for scalability, on which we layer higher-level APIs such as a POSIX file system and a Kafka-compatible message bus.   Cascade's native key-value (K/V) APIs are compatible with dataset APIs offered by mainstream systems like PyTorch, TensorFlow, and Spark.  Moreover, Cascade is designed to coexist with standard cloud tools.  The most significant benefits are for code that is able to run entirely within Cascade, but the design enables a very high degree of legacy compatibility.

Our main focus is on sophisticated AI applications structured as a pipeline or data-flow graph (DFG) of ML stages.  The experiment shown in Fig.~\ref{fig:kafka-direct} illustrates the challenge and opportunity.  We created a two-stage pipeline in which the ML stages are no-ops.  The blue latency distribution is from a pure-Cascade deployment, while the orange one runs the identical code on Kafka-Direct, a recent system that sets pub/sub records by leveraging RDMA~\cite{kafkadirect,kreps2011kafka}.  Here, tail latencies exceed 1 second: orders of magnitude worse.  As explained more fully in Sec.~\ref{sec:kafka-direct}, the right-hand figure shows that these high latencies arise in the communication infrastructure, not the platform.  The finding is not unusual:  today's cloud platforms achieve amazingly high throughput, but tail latency is a recognized problem (see, for example, the performance benchmarks in ~\cite{kreps2011kafka,scribe,googlepubsub,awskinesis, structured}).  Cascade shows that we can have low latency without loss of throughput.

Our contributions can be summarized as follows:
\vspace{-6pt}
\begin{enumerate*}
\item We have created a new open-source platform for hosting delay-sensitive MLs and the data on which they depend, supporting many popular ML backends.
\item Cascade offers multistage MLs a way to create {\em fast paths} between components that efficiently move data while minimizing delays.  Central to this is the avoidance of unnecessary data movement and copying.
\item We undertake an extensive evaluation that quantifies Cascade's costs and compares real-world applications hosted on Cascade with state-of-the-art alternatives.  Cascade reduces latency (sometimes by orders of magnitude) while sustaining high throughput.
\end{enumerate*}

\section{Architectural Considerations}
Four observations shape the Cascade architecture:
\\\\
{\bf MLs are graphs:}
Edge intelligence applications will generally be structured as mixtures-of-models (MOMs): pipelines or directed graphs.  An application might segment an image in a first stage, tag objects in a second stage, then gate further actions on these tags.  Even generative AIs adopt this approach: a so-called mixture-of-experts (MOE). Rather than optimizing edge platforms for a single program running on a pool of servers, we should focus on event-triggered graphical scenarios.  The MLs will often run in containers, and a MOM or MOE might combine programs coded in entirely different frameworks (PyTorch, Tensor Flow, Apache Spark, etc).  
\\\\
{\bf MLs must be tightly coupled to an O(1) data store:}  Large MLs are trained offline (on ``stale'' data), hence edge applications seeking a contextualized response must supplement the ML with ``world state'' data.  To some extent this is done with ``prompt engineering", but the more general case holds contextual data in tables and other objects, as in~\cite{memgpt}: the ML might directly consult these tables, or it could produce a generic response, then finalize it using contextual information.  Full-featured database support is unnecessary: ML languages allow embedded queries on collections so fast K/V lookups and updates suffice.
 \\\\
{\bf Collocating data and compute enables big wins:}
Latency-sensitive tasks are at risk of stalling when fetching objects from storage.  This is a particularly serious issue with intelligent applications: ML hyperparameters and models are often large objects, and copying them over a network will be costly even if the network is fast, but it could also arise when consulting contextualization data.  Collocating computational tasks with required data avoids these stalls.
\\\\
{\bf Accelerators impose special requirements:}  
Edge clusters will often have RDMA networking~\cite{simpson2020securing,tsai2019pythia,edgecloud}, together with GPUs and perhaps other hardware accelerators.  But these technologies impose requirements on the application.  An ideal edge solution must have an end-to-end zero-copy design, minimize locking, and allocate objects in cache-aligned, DMA-mapped, pinned pages~\cite{FaSST,rdmakafka,lu2014,rdmastorm,xue2019}.  This, however, can be at odds with the use of ML languages that often were designed before such considerations became important.  Many systems work around such issues by leaving legacy logic untouched (including legacy memory management) and then doing last-moment copying into properly prepared memory regions.  Our work identifies ways to run legacy code efficiently but without last-moment copying.

\section{Design and Implementation}\label{sec:design}

Leveraging the insights above, we created Cascade: an ML hosting platform that combines storage and compute to provide both low latency and high throughput, without compromising ease of development and legacy compatibility.  The core of the system is built in C++17 using an RDMA cloud library called Derecho~\cite{Jha2019} for fault-tolerant group communication.  Cascade has three main elements: (i) a {\em lambda} API for hosted application logic; (ii) a sharded K/V object store; and (iii) the {\em fast path} mechanism, which is a series of optimizations that glues together the other two elements while cutting overheads in the end-to-end data/compute path.  Each of these elements is described next, along with implementation details and other optimizations.

\subsection{The Lambda API: Developing Applications}
\label{sec:appabs}

An AI application on Cascade consists of a series of ML stages connected
to form a Data-Flow Graph (DFG).  A DFG vertex represents the computation of an ML
stage wrapped in a {\em lambda}, which consists of a piece of code that
processes input data objects and produces results as objects.  Directed edges
correspond to the flow of objects between lambdas.  Lambdas can be simple
programs,  distributed programs that leverage parallelism, or even complex ML
models that perform some tasks using special hardware.  Lambdas often
will be build using existing ML packages, like PyTorch. Developers provide them to Cascade
using the {\em lambda API,} which supports lambdas compiled to run as methods defined in Dynamically Linked Libraries (DLLs), or as normal programs encapsulated in containers.

As further described in section~\ref{sec:fastpath}, lambdas are hosted on the same servers (and potentially in the same memory address space) as the K/V store. However, this is transparent to developers: Cascade provides simple APIs for a lambda to interact with the K/V store -- Cascade SDK API, Cascade Kafka API, and Cascade FUSE file system.  Lambdas use one of these APIs to retrieve objects required for computation, and to store results.  Currently, Cascade provides APIs for developing lambdas in C++, Python, and C\#. 

Porting an existing ML application to Cascade is a trivial procedure.  A JSON file describing the application DFG must be uploaded to Cascade.  Then, for each lambda, a developer must create a thin wrapper using the lambda API.  This wrapper has two responsibilities: (i) to provide Cascade with an upcallable function; and (ii) to retrieve input objects from and store output objects into the K/V store using one of the provided APIs.  An example is described in section~\ref{sec:rcd_app}: we deployed the same application both on Cascade and on Microsoft Azure. The effort to deploy our ML models on Cascade ($\sim$200 lines of Python code per model) was on par with commercial offerings such as Azure Machine Learning.

\subsection{K/V Store and Data Organization}
\label{sec:dataorg}

Cascade runs as a distributed, scalable service on a cluster of nodes, each
capable of hosting data and performing computation.  The native storage
abstraction is that of a sharded, fault-tolerant, versioned K/V object store.
Keys map to shards to determine the {\em home shard} of an object, using a
hashing scheme that can be customized to group objects so that related objects
will always be hosted on the same nodes. Objects are retrieved through {\em get}
operations, and stored through {\em put} operations. Tables can be stored as single
objects or spread over the store, using the unique row ID as part of the key.

Objects are managed in {\em pools}, each identified by
a path prefix.  Each object pool has a an access control policy, a replication factor, persistence/logging
properties, and a sharding policy. Object keys are prefixed by the pool path,
and then can have any desired path suffix. Note that although Cascade pathnames
permit a hierarchical organization of pools, any given object resides in just a
single pool.

Cascade offers three levels of data persistence.  Transient objects are used to
initiate a lambda computation but then vanish. In the smart traffic intersection example we present in section ~\ref{sec:rcd_app},
raw images arrive as transient objects.
A {\em volatile} pool retains the latest
version of objects in main memory for performance and short-term usage.  The smart intersection uses this mode to record trajectories of active objects.
A {\em persistent} pool logs versions, holding the latest data version in memory  but also
retaining all versions on persistent storage.  This permits versioned or
time-indexed data retrieval (as in ~\cite{song2016}). Cascade automatically timestamps
 objects, making it easy to track the evolution of some phenomenon over
time, hence a persistent object pool could be used to log 
sequences of photos and trajectories leading to an accident.
Performance optimizations used to facilitate persisted log access are described in
Section~\ref{sec:persopt}.

A trigger put sends an object to 
a randomly selected member of the target object pool using a Derecho P2P send; if a lambda is
watching for this object on that member, computation will be initiated, receiving the K/V pair as its arguments.  A put to a
volatile pool is implemented as a Derecho atomic multicast. A put to a persistent pool maps to a
Derecho replicated log update (a version of Paxos that updates 
all log instances, not just a quorum subset). 

Notice that both forms of put maintain 
identical states in all shard members, hence a get can be randomly sent to any 
member of an object's home shard, which will reply with the desired version.

Importantly, we run trigger put and general put on a different thread than the
one used to implement get operations.  This improves concurrency, but 
objects can be overwritten by simply
issuing a put using the same key, resulting in a chain of versions.
Howover, a get could occur just as an update is occurring.
We avoid the need for locks by using two atomic version numbers for
each object, $v_a$ and $v_b$.  The put operation updates $v_a$, then updates the
object data, and finally updates $v_b$. A get accesses $v_b$, then reads the object data,  then compares $v_b$
against $v_a$. If $v_a$ and $v_b$ differ, the get is reissued. 



\subsection{Fast Path: Where Data Meets Computation}
\label{sec:fastpath}

Cascade's fast path is a heavily optimized critical path enabling end-to-end data flows and computation 
to run at the highest possible speed: a {\em cascade} of actions.  It is used only for collocated computation:
 lambdas run on the servers where Cascade itself is running. In what follows we
first describe the fast path for a case in which all the lambdas are directly callable functions implemented
by DLLs that Cascade loads on startup.  Then we show how
this same scheme can be emulated to support containerized lambdas.  Lambdas running in the Cascade address space must be trusted,
hence although this case is slightly faster, it cannot be used with potentially malicious code.

Consider an incoming K/V trigger put or put on the Cascade system thread.  
The question arises of how to implement rapid lambda upcalls.
We rejected the idea of just doing direct upcalls from Cascade's system thread:
delays could disrupt the entire system.  
A second option we rejected introduces a
 {\em dispatcher} designed to fork off threads to run the triggered lambdas, similar to the
event loop in a web server.  It turns out that such a design runs into a number of
issues.  Thrashing could occur if the dispatcher were to launch too many threads
concurrently.  Delay-sensitive applications would incur the cost of thread
forking and frequent context switches.  Moreover, because one incoming object
could match multiple path prefixes and trigger multiple lambdas, such a design
results in a complicated interaction between the dispatcher and the per-lambda
execution threads. 
\begin{figure}[t]
 \centering
 \includegraphics[width=\linewidth,,] {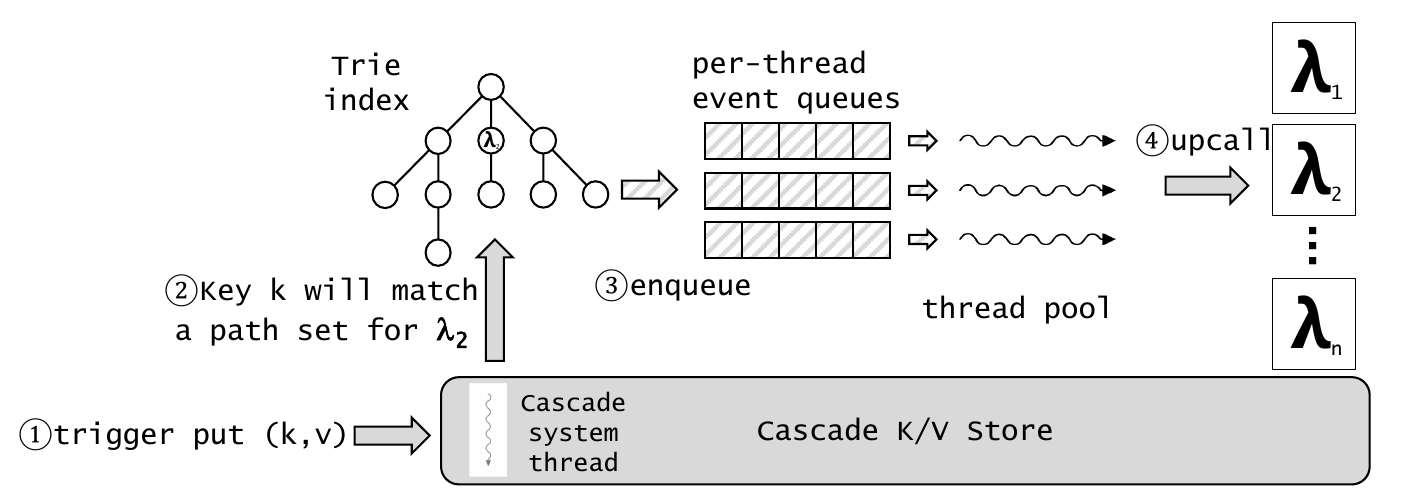}
    \caption{Fast path in a Cascade Node}
 \label{fig:fastpath}
\end{figure}

Accordingly, we settled on the fast path architecture seen in Fig.~\ref{fig:fastpath}.  This still uses a dispatcher thread that interacts with the Cascade core event loop; it will observe a stream of K/V updates as Cascade receives them (\circled{1}).  However, its only role is to detect path prefix matches (\circled{2}) and, for each match, to enqueue an {\em upcall event} containing a pair of shared pointers to the object and the matching lambda (\circled{3}).  Path matching is accelerated by using a {\em trie} data structure.  To the right, we maintain a small pool of lambda upcall threads.  Each of these threads loops, dequeuing incoming upcall requests and calling the appropriate lambda (\circled{4}).

Notice that each upcall thread has its own event queue.  By default, the dispatcher enqueues in round-robin order for load-balancing.  However, if
a lambda is configured (in the DFG) for FIFO invocations, the dispatcher picks a queue based on the key hash of the lambda's object so that objects sharing the same key (for example, the video frames from the same camera) will occur on the same thread.

For containerized lambdas running on a server where Cascade it also running, we mimic the DLL model using IPC: the API is the same, but the lambda upcall occurs in a container with its own address space on the same machine as the Cascade node.  Data is moved between the container and Cascade using a pair of shared-memory segments.  One segment is limited to read-only access, and lets Cascade selectively reveals hosted data to the application.  The other segment lets the application send data to Cascade; it additionally holds two lock-free FIFO queues, one used for fast IPC and the other for responses and notifications.  In our experiments, DLL hosting is slightly faster, but it is obviously less secure.  The containerized approach benefits from better security, but system calls are slightly more costly:  an RPC though our shared memory FIFOs incurs round-trip latencies varying from about 0.5$\mu s$ to slightly more than 1.5$\mu s$, depending on the server speed. 

\subsection{Data Movement}
Throughout Cascade, the need to pass objects from place to place could have caused a great deal of copying.  Common reasons include
 data marshalling, the need to gather data scattered over multiple memory locations, and cacheline or page alignment needs~\cite{JRB21}.   To minimize copying, we needed to minimize costly data layouts.



An example arises with in-place C++ object construction.  Here the classic reason for copying is to address hardware-specific byte ordering and alignment.  Our approach assumes that servers will all use the same native data layout rules.  External clients respect server policies even if they differ from the native ones, potentially paying a cost similar to standard marshalling if they run on hardware that favors some other native data formatting. This enables RDMA transfers without marshalling or copying.  Cascade publishes the service's desired byte layout and alignment when an external client connects, enabling clients to preconstruct efficient marshallers for native types at binding time.   For small amounts of data, the costs will be negligible. Large objects like photos or video often have architecture-independent encodings; here, the main need is to ensure that when such an object is first uploaded by the device, it will be saved to a suitable memory region.  Jointly, such steps eliminate marshalling for most objects (marshalling cannot be avoided for complex structures or ones that contain pointers, such as lists).  

\subsection{Resource scheduling}

As noted earlier, Cascade uses a deterministic hash-based  policy to map objects to shards.  Then, to promote load-balancing, we use a round-robin policy to decide which node in a shard should process each matching input object.  The effect is that tasks run on nodes that already have the required ML model: often (by far) the largest data dependency.  Cascade then uses an LRU cache to retain additional objects that the computation happens to access: after a short warm-up, all members in a shard will hold copies of any systematically-required data object.  The system additionally supports customization of these policies, leaving space for more sophisticated scheduling and data collocation optimizations.  Other control knobs include CPU-core-to-thread-pool mapping and NUMA node control.  The experiments reported here all use the defaults for these settings.


\subsection{Persistent Log Optimizations}
\label{sec:persopt}
We employ three methods to accelerate operations on the persisted log mentioned in Section~\ref{sec:dataorg}.  First, log files are memory-mapped, enabling a simplified implementation of log read/write.  Next, a write-back thread asynchronously flushes the opportunistically batched updates to the filesystem for efficient write performance.  Finally, we link logged records associated with the same key using backpointers~\cite{balakrishnan2013tango} to accelerate range queries and temporal indexing.

A quick example will illustrate how these features are employed.  When a put is performed in a persistent object pool, for example to store a new image, Cascade uses Derecho's Paxos-based logging protocol to atomically update an in-memory copy of the object, while also linking the new version to the prior one using meta-data Cascade caches in memory for all active objects. As a result, the Derecho protocol only needs to perform a single disk write.

Once the data is stable in the log, Cascade extends its indexing data structures by adding a new entry for the new version, and then updating a temporally-sorted secondary index so that time-indexed object retrieval can be mapped to versioned object retrieval.  This uses a standard C++ sorted queue, and is very fast compared to the cost of disk I/O, hence it has a negligible impact on performance of the put operation. At this point, Cascade can permit access to the stable prefix of the log: although there may be a pending suffix that is still being updated, updates to already logged records are not permitted.  The data structures used are very similar to those described in ~\cite{song2016}.

For a get request, there are three cases.  A get that simply requests the most current data will receive the currently stable in-memory version of the desired object.  A get requesting a specific version or a range of versions is satisfied by scanning the linked version chain to extract a series of pointers.  A temporal get range query is resolved by first mapping the time window to a version window, and then using the same logic as for a versioned range query.  Should a time window extend beyond the stability frontier of the system (``into the future''), we transparently delay, satisfying it only when the desired time range is fully stable.  This avoids the risk that the get might return a result omitting some version that the window should have included.

\section{Evaluation}
\label{sec:eval}

In this section, we first present microbenchmarks that evaluate latency and throughput of the Cascade K/V store. We then evaluate the performance of the fast path using a data pipeline composed of multiple put operations, employing no-op actions to highlight overheads. Later, in Section~\ref{sec:applications}, we consider real applications.

Our computing environment consists of dedicated servers equipped with Mellanox ConnectX-4 VPI NIC cards connected to a Mellanox SB7700 InfiniBand switch, which provides an RDMA-capable 100Gbps network backbone.  The servers have two configurations.  The more powerful configuration (denoted {\bf type A}) matches what edge-hosting systems typically offer; these have dual Intel Xeon Gold 6242 processors, 192 gigabytes of memory, and an NVIDIA Tesla T4 GPU.  The lightweight setup (denoted {\bf type B}) emulates a less powerful Cascade client; it has two Intel Xeon E2690 v0 processors and 96 gigabytes of memory but no GPU. Both types of servers have high-speed NVMe storage (Intel Optane P4800X cards).
All servers run Linux kernel 5.4.0.  We synchronize the server clocks with PTP~\cite{ptp}, allowing comparison of timestamps from different servers with sub-millisecond precision.  

\subsection{Cascade K/V Store Performance}

\begin{table}
  \small
  \begin{center}
    \begin{tabular}{l|l|l|l|l|l|l}
      \hline
      \multirow{2}{2.2em}{Object Size} & \multicolumn{3}{c|}{Put Latency ($\mu s$)} & \multicolumn{3}{c}{Get Latency ($\mu s$)} \\
      \cline{2-7}
      & \texttt{trig} & \texttt{vola} & \texttt{pers} & \texttt{-100ms} & \texttt{-1s} & \texttt{-10s} \\
      \hline
      10KB & 12 & 70 & 500 & 67 & 68 & 67 \\
      \hline
      1MB & 220 & 1100 & 4200 & 1304 & 2200 & 2094 \\
      \hline
    \end{tabular}
    \caption{Typical K/V Store Access Latencies}
    \label{tbl:kv_lat}
  \end{center}
\end{table}

\begin{figure}[]
  \centering
  \subfloat[10KB object\label{fig:put_thp_10k}]{
    \includegraphics[width=0.46\linewidth,,]{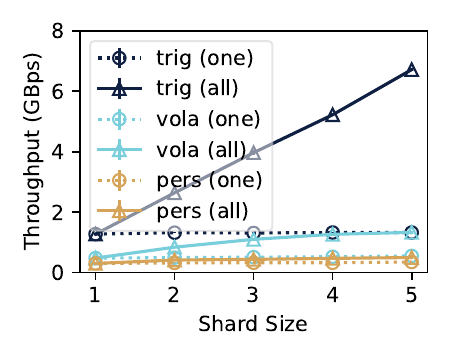}
  }
  \subfloat[1MB object\label{fig:put_thp_1m}]{
    \includegraphics[width=0.46\linewidth,,]{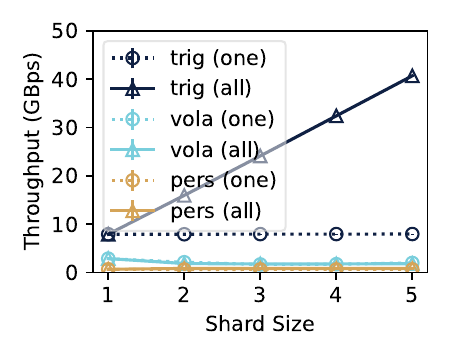}
  }
  \caption{K/V Store Put Throughput}
  \label{fig:put_thp}
\end{figure}

\begin{figure*}[]
  \centering
  \subfloat[Trigger put with 10KB objects\label{fig:tput_lat_10k}]{
    \includegraphics[width=0.31\linewidth,,]{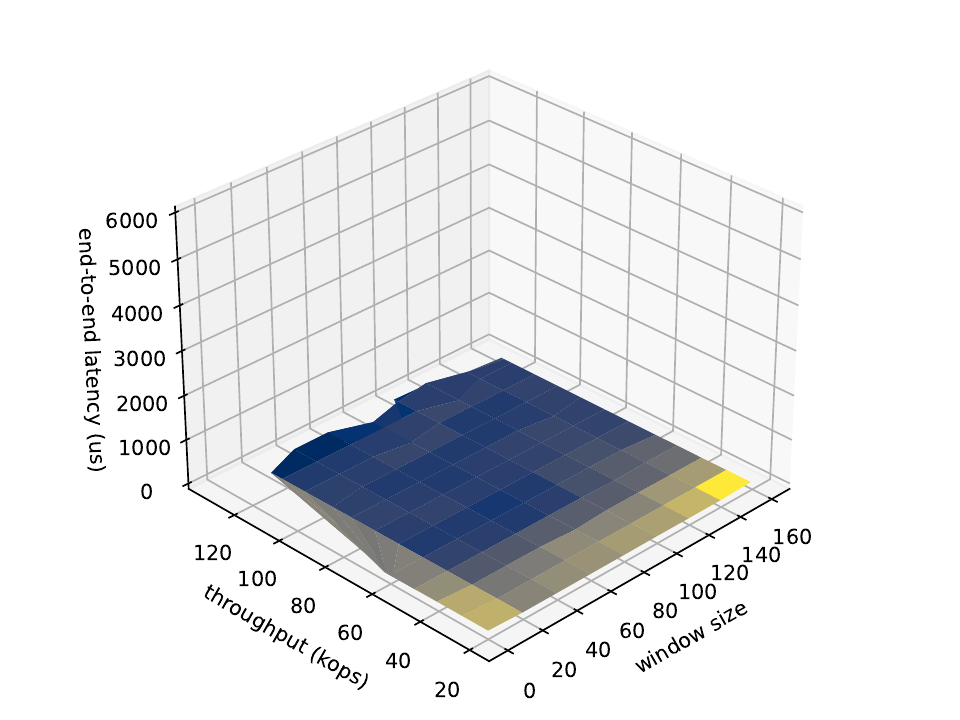}
  }
  \subfloat[Volatile put with 10KB objects\label{fig:vput_lat_10k}]{
    \includegraphics[width=0.31\linewidth,,]{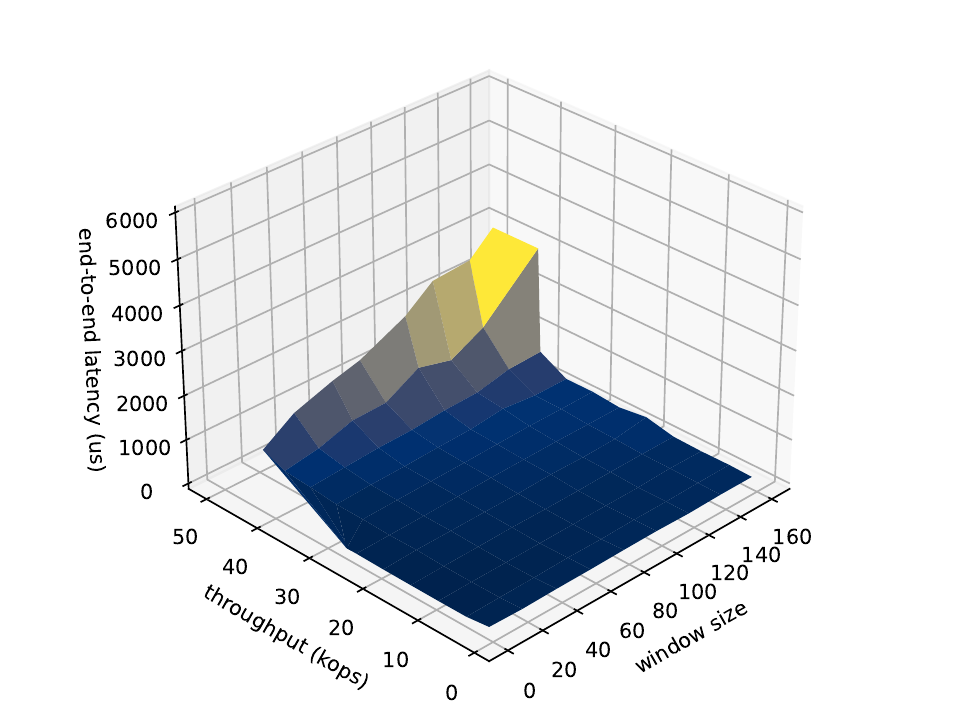}
  }
  \subfloat[Persistent put with 10KB objects\label{fig:pput_lat_10k}]{
    \includegraphics[width=0.31\linewidth,,]{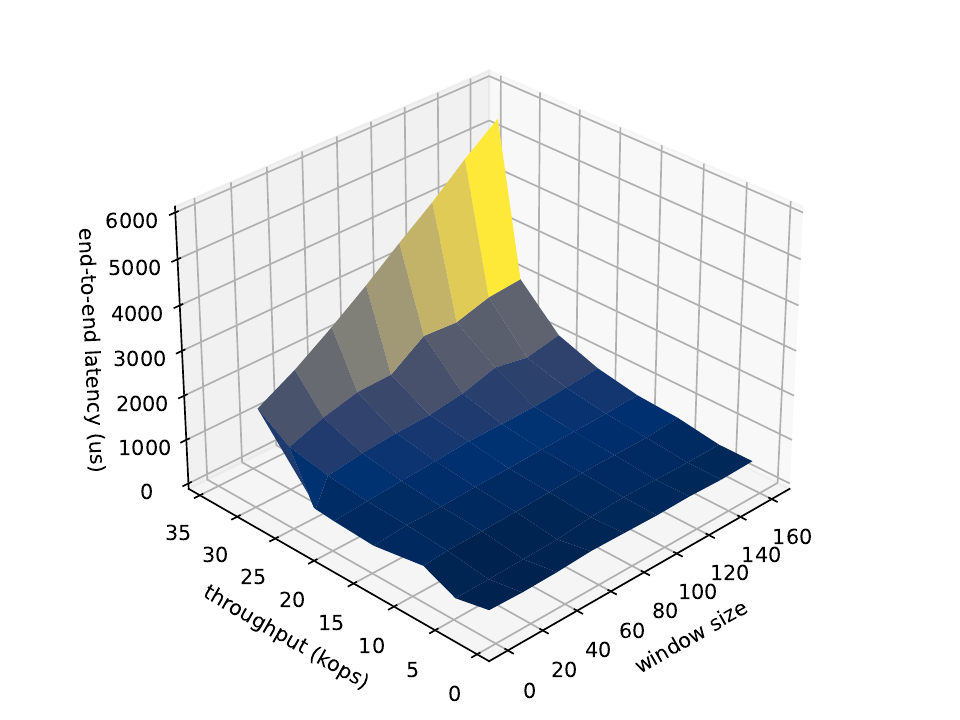}
  } \\
  \subfloat[Trigger put with 1MB objects\label{fig:tput_lat_1m}]{
    \includegraphics[width=0.31\linewidth,,]{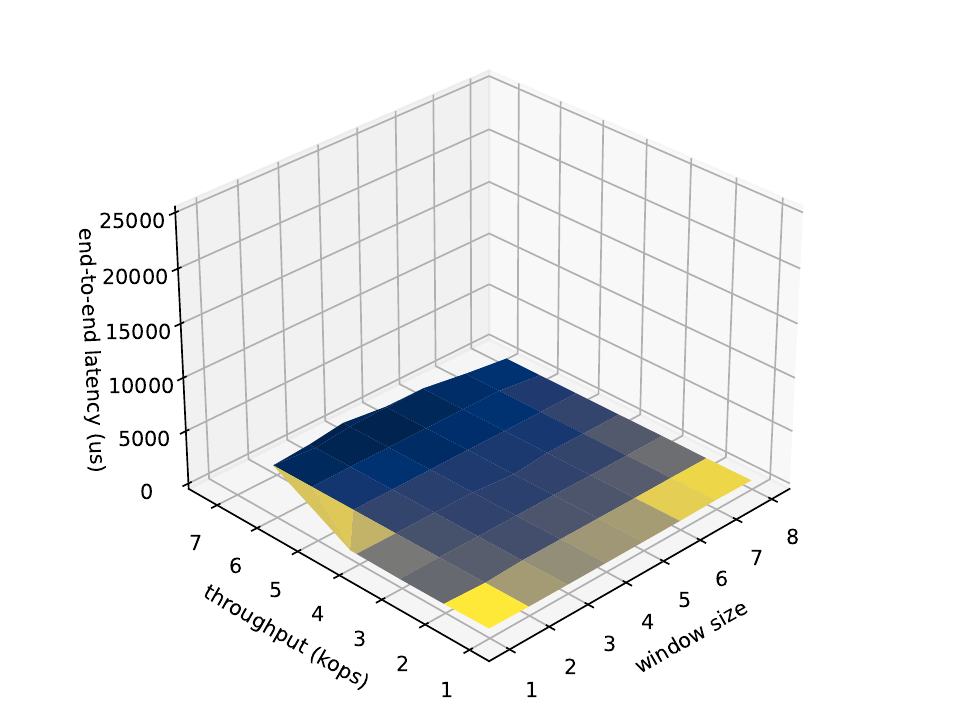}
  }
  \subfloat[Volatile put with 1MB objects\label{fig:vput_lat_1m}]{
    \includegraphics[width=0.31\linewidth,,]{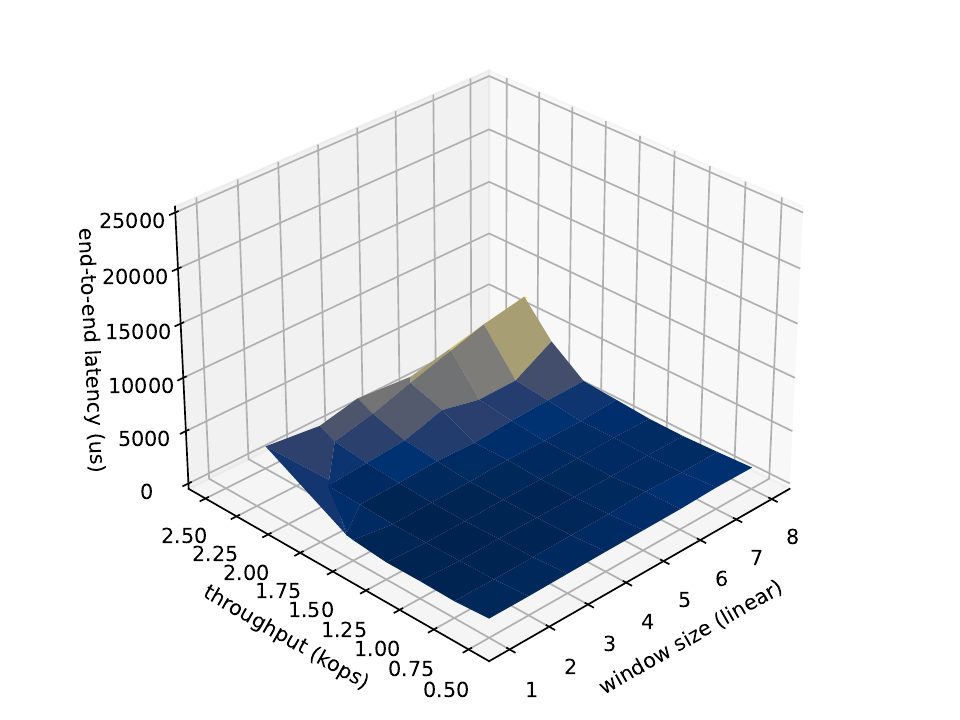}
  }
  \subfloat[Persistent put with 1MB objects\label{fig:pput_lat_1m}]{
    \includegraphics[width=0.31\linewidth,,]{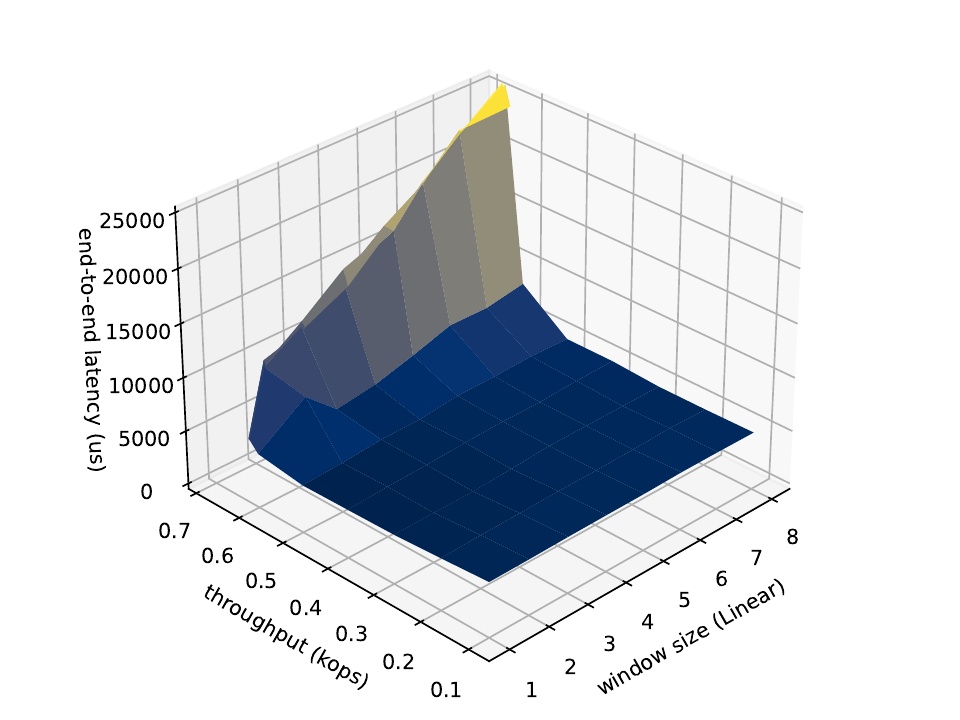}
  }
  \caption{Cascade's K/V Store Put Latency is low over a wide range of object sizes and data rates.}
  \label{fig:put_lat}
\end{figure*}

\begin{figure*}[]
  \centering
  \subfloat[Volatile put\label{fig:vput_breakdown}]{
    \includegraphics[width=0.49\linewidth,,]{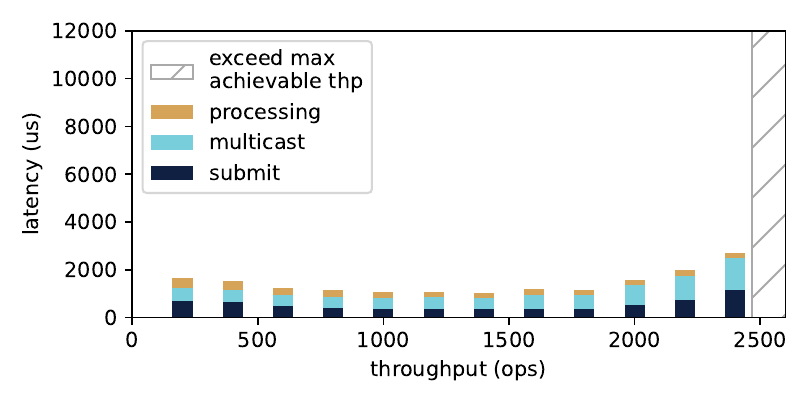}
  }
  \subfloat[Persistent put\label{fig:pput_breakdown}]{
    \includegraphics[width=0.49\linewidth,,]{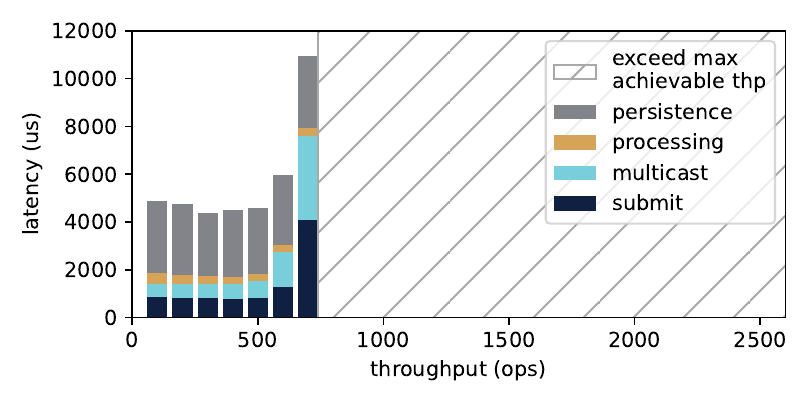}
  }
  \caption{Latency Breakdown for Volatile and Persistent Put}
  \label{fig:put_lat_breakdown}
\end{figure*}

We present now experimental results on the performance and scalability of Cascade's put and get operations.
The three types of put provided by Cascade (described in section~\ref{sec:dataorg}) are denoted as \texttt{vola} (volatile), \texttt{pers} (persistent), and \texttt{trig} (trigger).
We evaluated the time-indexed version of the get operation on a persistent pool in which test objects had new versions every 10 ms, and issued get requests for timestamps at varying distances from the current time.

Clients run on type B servers, issuing requests to Cascade nodes on type A servers at a controlled rate.
To evaluate scalability, we varied the shard size as well as the number of clients. Some experiments employed just a single client irrespective of shard size (\texttt{one}), while others had multiple clients, one per shard member (\texttt{all}).


Table~\ref{tbl:kv_lat} explores the latency of above operations considering small and large objects.
For each \texttt{vola} put operation, we measured time starting when the client sends the request, and ending when all replicas finish updating their in-memory store.
Each data point in the figure shows the average latency during a five-second period.
Similarly, for each \texttt{pers} put operation, we measure time from when the client first sends the request to when the last replica finishes persisting it.  For {\tt trig} put we measure until the request reaches a replica that upcalls to a developer-supplied lambda.  

The \texttt{pers} put latency is about four to five times higher than that of a \texttt{vola} put.
We confirm that the bottleneck is the I/O to our storage devices.
{\tt Trig} put is one order of magnitude faster than volatile and persistent put because it does not need to replicate or persist any data.

The latency for a time-indexed get is comparable to that of a \texttt{vola} put for small objects, and approximately twice as long for large objects.
The overhead comes from the extra step of sending the data back to the client after the server receives and handles the request.
The results do not show significant performance differences between fresher and staler data: our versioned log-searching operations are efficient, and the page cache will hold the requested section of the log after the first access.

The data in Table~\ref{tbl:kv_lat} reflects performance when the system is not saturated.
As the workload is increased and begins to approach the maximum sustainable throughput, latency will rise sharply and without limit.  To quantify this effect, we measured the end-to-end latency with varying object sizes and rates as shown in Fig.~\ref{fig:put_lat}. The six subfigures show the latency for the three operation types and two different object sizes.  Again, we use three replicas for \texttt{vola} and \texttt{pers} put operations.

As shown in figures~\ref{fig:vput_lat_10k}, \ref{fig:pput_lat_10k}, \ref{fig:vput_lat_1m}, and~\ref{fig:pput_lat_1m}, before the system becomes saturated end-to-end latency is consistently low, corresponding to the flat part on the near right part of the curved surfaces.  The system keeps up with the request rate and no queuing backlogs arise.  As the workload grows we see a sharp rise in latency, processing becomes bursty and queuing delay dominates the end-to-end latency.  Here, Cascade (and Derecho, which is moving the data) have started to batch in an opportunistic way, noticing small backlogs and then handling them as an ad-hoc mini-batch.  The system keeps up with the incoming load, but is no longer achieving low latency for requests that ended up on the queue for more than a few microseconds.

Fig.s~\ref{fig:tput_lat_10k} and \ref{fig:tput_lat_1m} show that the {\tt trig} put latency is insensitive to workload and window size.  This is a consequence of using a no-op as the triggered action:  If we used a lambda that performed a more realistic computation, the computing cost would dominate the end-to-end latency.

Fig.~\ref{fig:put_lat_breakdown} shows the latency breakdown for the {\tt vola} and {\tt pers} put.
We use a setup with a shard of three nodes and a client that uploads 1MB objects.
The window size is three.
The {\em submitting} component refers to the latency between the client serializing a put request into the sending buffer and the Cascade server receiving it;
the {\em multicast} component is the latency of replicating the data among the shard members;
the {\em processing} component is the time spent in updating the in-memory state,
and the {\em persistence} component represents the time between initiation of an update and its persisted commit (across all replicas).
Because {\tt trig} put has only the submitting component, we exclude it in this figure.

All options have a wide range of object sizes and data rates for which per-event latency remains stable and very low.   For {\tt vola} put, the multicast and submitting components account for most of the end-to-end latency.
For {\tt pers} put, the persistence overhead dominates.
When the request rate approaches the maximal achievable single-event throughput, the overhead of submitting and multicast grows suddenly because the messages pile up.  In the persistent case the effect is exacerbated by a stage in Paxos that syncs data to storage.


Fig.~\ref{fig:put_thp} shows the put throughput of Cascade K/V store as we vary the shard size (the number of replicas).
The Y-axes represent the throughput seen by the application.
With only one node in a shard, volatile put reaches $\sim500$ MBps (50 kops) and $\sim2.8$ GBps (2.8 kops) for 10KB and 1MB objects.
The throughput for 10KB objects is steady as we vary the shard size from 1 to 5.
With 1MB objects, both figures drop slightly: $\sim2.2$ GBps (2.2kops) for shards of size 5, reflecting the overhead of replicating large objects.

With multiple clients, we achieve even higher throughput.  Volatile put with 10KB objects rises from $\sim500$ MBps (25 kops) to $\sim1.3$ GBps (130 kops),
a figure at which the replication capacity of the system becomes saturated.
In contrast, with 1MB objects, even with multiple clients, throughput remains flat, peaking at $\sim2.7$ GBps for 5-member shards.
Our studies suggest that the bottleneck is associated with a {\tt memcpy} operation that we use to copy data from the RDMA buffers used for incoming data to a heap where we
store objects that will be passed to lambda upcalls.
In future work we plan to explore a redesign that would RDMA into per-message buffers within the heap.

The numbers for persistent put operations show similar trends, but the actual bandwidths are sharply reduced.
Persistent put reaches at most $\sim270$ MBps (27 kops) for small objects and $\sim800$ MBps (800 ops) for large ones.  The bottleneck turns out to be a side-effect of
the strong consistency model used in Derecho.  
Although our NVMe device can achieve sequential write bandwidth of 2.4 GBps, this data rate is only achievable with long, uninterrupted DMA transfers.  It turns out that in the persisted mode our update workload incorporates ordering dependencies that the storage layer enforces by periodically pausing until persisted updates are completed, disrupting the DMA transfer scheme (we plan to look at ways of aggregating such actions opportunistically, but this is future work).
{\tt Trig} put operations scale best because these operations avoid all memory copying and replication overheads.
A {\tt trig} put client gets $\sim7.6$GBps ($\sim7.6$ kops) for 1MB objects, which is close to the RDMA hardware limit, and aggregated throughput grows linearly in the number of clients.

\subsection{Fast Path Performance}

\begin{figure}[]
 \centering
 \subfloat[10KB Object\label{fig:fpbd10k}]{
   \includegraphics[width=0.49\linewidth,,]{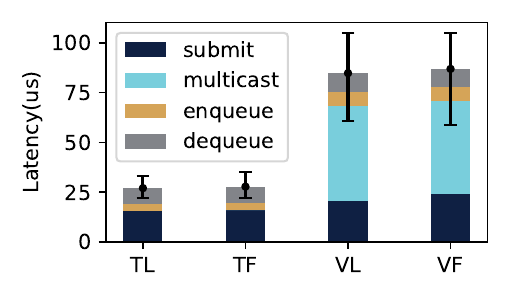}
 }
 \subfloat[1MB Object\label{fig:fpbd1m}]{
   \includegraphics[width=0.49\linewidth,,]{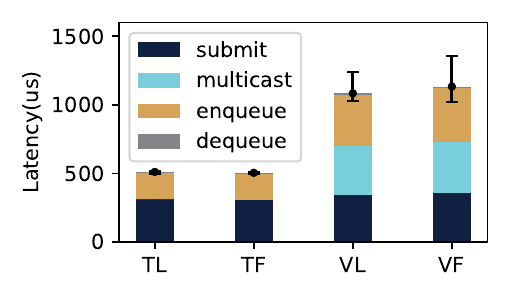}
 }
 \caption{Fast Path Latency Breakdown. T: trigger put; V: put with volatile persistence; L: load balancing; F: FIFO.}
 \label{fig:fpbd}
\end{figure}

\begin{figure}[]
 \centering
 \subfloat[Latency (10KB)\label{fig:pp_lat_10k}]{
   \includegraphics[width=0.33\linewidth,,]{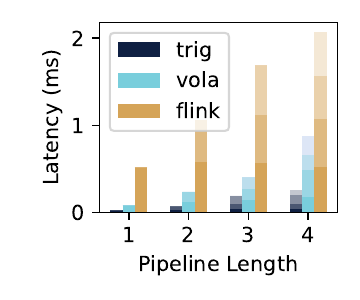}
 }
 \subfloat[Latency (1MB)\label{fig:pp_lat_1m}]{
   \includegraphics[width=0.33\linewidth,,]{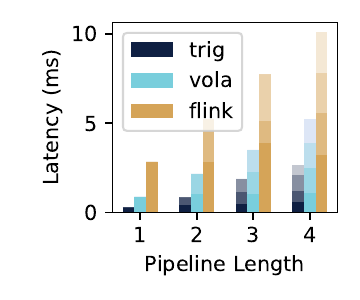}
 }
 \subfloat[Throughput\label{fig:pp_thp}]{
   \includegraphics[width=0.33\linewidth,,]{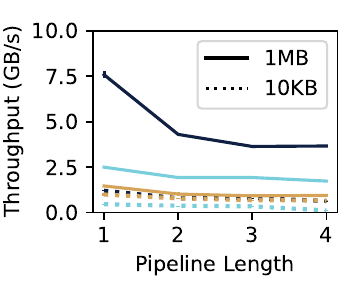}
 }
 \caption{No-op Pipeline Performance}
 \label{fig:ppl}
\end{figure}

We now evaluate the performance of the fast path.
Fig.~\ref{fig:fpbd} breaks down the fast path latency from the moment an object is sent to when it is being processed by the lambda on the next-hop node.
The {\em submit} component refers to the latency for the object data being transfered to the next hop (step \circled{1} in Fig.~\ref{fig:fastpath}).
The {\em multicast} component gives the latency of replication among shard members.
Trigger put does not have such a component as it does not replicate the data.
The {\em enqueue} component involves the dispatcher matching the object key with lambdas and inserting an event into a queue (steps \circled{2} and \circled{3}).
Our experiment shows that trie matching spends only $\sim 130 ns$ per depth level using a full ternary tree.
Here we use a key path with four levels.
The {\em dequeue} component involves an upcall thread picking the event and calling the lambda (step \circled{4}).

Fig.~\ref{fig:fpbd10k} shows the fast path latency breakdown for 10 KB objects.
The end-to-end latency is $\sim25 \mu s$ for trigger put (the {\em T} prefix) and $\sim85 \mu s$ for put with three replicas in volatile pool (the {\em V} prefix).
Compared to the corresponding put latency, the $\sim15 \mu s$ overhead shows that the fast path is lightweight.
We also compared the load balancing (the {\em L} suffix) and FIFO (the {\em F} suffix) dispatch policies.
Results show no significant differences between them.
For 1MB objects, Fig.~\ref{fig:fpbd1m} shows a similar trend.
However, the {\em enqueue} component is more significant than that of small objects, because copying is more costly even though we only make one copy for multiple upcalls.
We plan to address it with a zero-copy solution in future work.

With growing date rates, we found that the latency was low and quite stable until the system becomes saturated.
Once the data rate reaches the limit for per-event upcalls, throughput stops growing and latency surges dramatically as requests
begin to wait on congested queues.
To stimulate typical latency behavior, we used 10K/1K ops data rate for 10KB/1MB objects respectively.

To microbenchmark a longer fast-path pipeline we created a series of lambdas that each relay received data but perform no other computation, implemented with a trigger put (\texttt{trig}) or a three-replica volatile put (\texttt{vola}).  Each lambda ran in an A-type server node.  A first experiment examined latency from the client to the no-op for 10KB and 1MB objects sent at a moderate request rate.  In Fig.~\ref{fig:pp_lat_10k} and \ref{fig:pp_lat_1m} we show the average latency during a representative 5-second period for varying pipeline lengths.  The single-stage numbers match the fast-path latencies in Fig.~\ref{fig:fpbd}.  The pipeline is faster with the trigger put than with the volatile put, but the overhead of volatile put is surprisingly low.

For purpose of comparison, we then configured Apache Flink to mimic Cascade by having it use a single task-processing slot per server, disabling automatic operator chaining to prevent it from merging the tasks.  We recompiled Flink to load 1MB at a time (the default is 32KB), and set its minibatch delay barrier to zero.  This last change goes beyond the norm but without it the performance was terrible and the servers had very low CPU utilization levels.  Yet even with all of this tuning, Flink's pipeline latency is high (\texttt{flink}).
The Cascade pipeline with trigger put has a latency below one-eighth that of the Flink version for 10 KByte messages and one-fourth for 1 MByte messages. Indeed, even the (replicated) volatile put on three-member shard has less than half of the latency of Flink, at both message sizes.  Two factors account for this: Flink runs on TCP, not RDMA, and it uses the 
Java-based Kryo serializer, which copies from Java-managed memory to a network buffer.    

Finally, we stressed each pipeline by streaming at the maximum sustainable message rates.  This yields the first four throughput series in Fig.~\ref{fig:pp_thp}.
Again, the throughput of Cascade's one-stage pipeline matches the trigger and volatile {\tt put} throughput in Fig.~\ref{fig:put_thp}, dropping slightly as we move to a pipeline with two or more stages.  This reflects the extra costs associated with message relaying: the first stage only sends, while inner pipeline stages must send and receive, and the final stage only receives.   Performance is sustained as pipeline length grows from two to four, supporting our claim that Cascade scales extremely well.
In the same experiment, Flink gives lower and more variable throughput.

\section{Cascade Applications}\label{sec:applications}

In this section, we evaluate three different applications on Cascade. The goal is twofold: (i) to show the performance benefits of Cascade when compared to alternative solutions; and (ii) to present evidence of Cascade ease-of-use, in particular when deploying off-the-shelf ML models. We made a major effort to be fair to the comparison platforms, and to configure them exactly as recommended. The computing environment is the same as in Section~\ref{sec:eval}.

\begin{figure}[t]
  \centering
  \subfloat[Messaging Service\label{fig:cms_dfg}]{
    \includegraphics[width=0.29\linewidth,,]{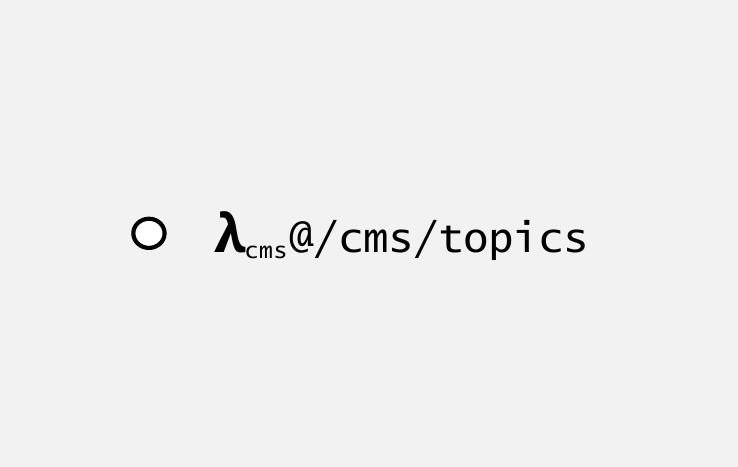}
  }
  \subfloat[Smart Farm\label{fig:sf_dfg}]{
    \includegraphics[width=0.31\linewidth,,]{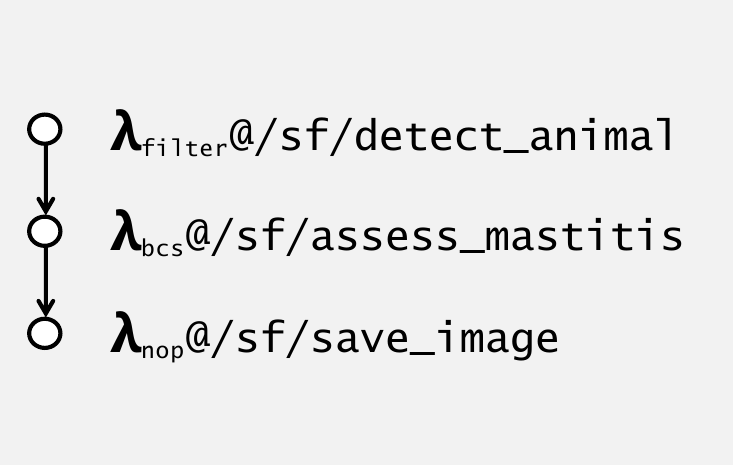}
  }
  \subfloat[Collision Detection\label{fig:rcd_dfg}]{
    \includegraphics[width=0.31\linewidth,,]{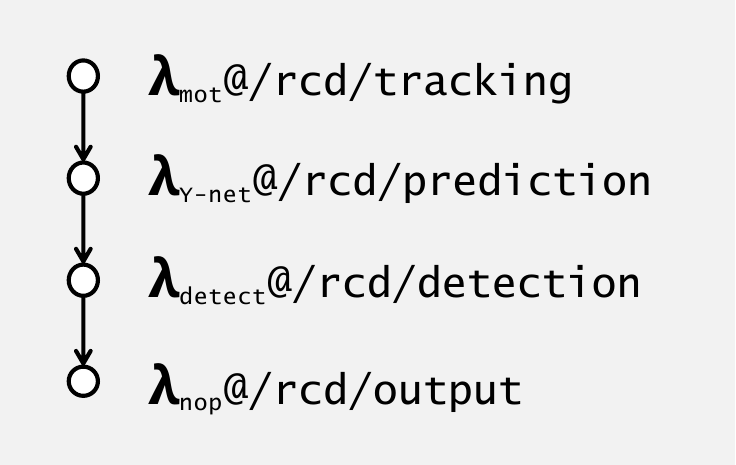}
  }
  \caption{Application DFGs}
  \label{fig:dfgs}
\end{figure}

\subsection{Cascade Messaging Service} \label{sec:kafka-direct}

Our first application is the {\em Cascade Messaging Service (CMS)} that was compared to Kafka-Direct in the introduction. The CMS employs a standard Pub/Sub model but maps publish to atomic multicast, enabling strong ordering and fault-tolerance semantics. 
As seen in Fig.~\ref{fig:cms_dfg}, the CMS DFG has just one vertex: a lambda ${\lambda}_{cms}$ binding to path {\em /cms/topics}. A CMS client publishes to a topic $T$ by calling {\tt put} with key {\em /cms/topics/T}, and will be either volatile or persistently logged at the developer's option. ${\lambda}_{cms}$ does no computation; instead, it pushes a notification (including the object data) to clients subscribed to $T$. ${\lambda}_{cms}$ is FIFO by topic, but allows concurrent upcalls for distinct topics. 

We hosted both platforms on type A servers, and configured Kafka-Direct as recommended by the developers, with publisher, subscriber and server nodes on distinct machines. To create Fig.~\ref{fig:kafka-direct} (left side) we disabled intentional batching to prioritize latency over throughput (nonetheless, both batch if a backlog forms).  To avoid bottlenecks in the persistent storage layer, $\lambda_{cms}$ uses a volatile pool, while KafkaDirect's log data was stored in ramdisk.   We then varied the data rates at each object size.  For each size we were able to identify rates that minimized latency for both systems.   We then plotted the median (circles), 10\%-90\% range (box) and an error bar for the 1\%-99\% range. The right side of the figure breaks delay down by the stage at which it arose.   

\begin{figure}[t]
  \centering
  \includegraphics[width=0.85\linewidth]{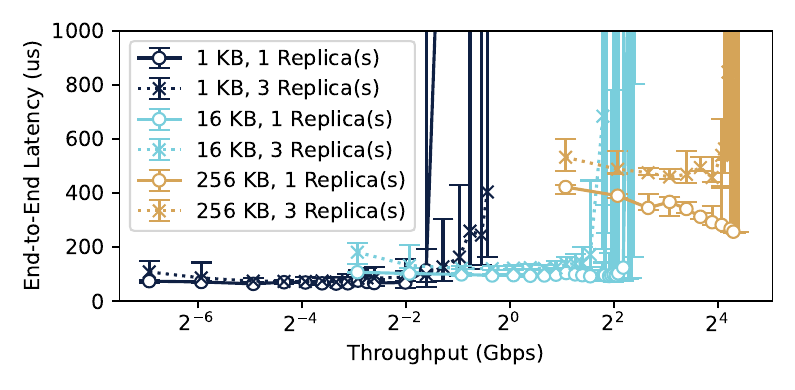}
  \caption{CMS Latency: Detailed Breakdown}
  \label{fig:cmslat}
\end{figure}

Fig.~\ref{fig:cmslat} shows latency at each presented data rate, but only for CMS; we see steady low latency as long as the system can keep up event-by-event, and then a shift to a batched behavior as backlogs begin to form.  We were unable to create a similar graph for Kafka-Direct: latencies fluctuated widely even at very low event rates.

\begin{figure}[t]
  \centering
  \subfloat[Latency Breakdown\label{fig:dflat}] {
    \includegraphics[width=0.48\linewidth,,]{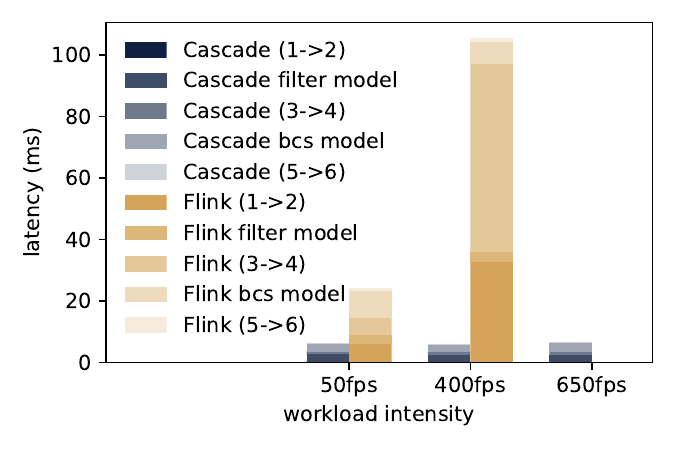}
  }
  \subfloat[Txns/sec versus stage size (front-end,compute)\label{fig:dfthp}] {
    \includegraphics[width=0.48\linewidth,,]{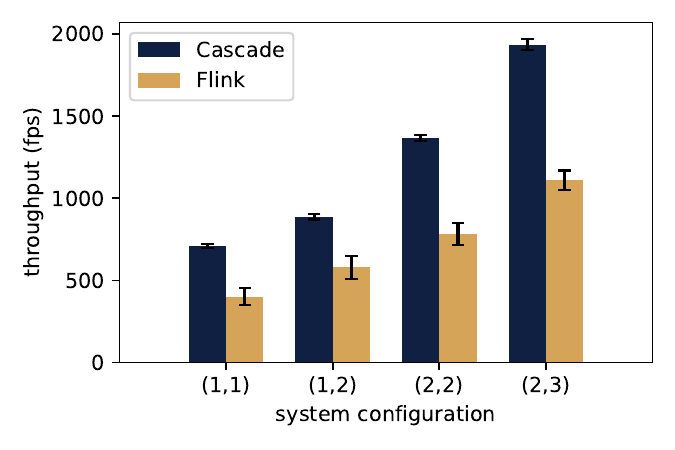}
  }
  \caption{Smart Farming Experiment}
  \label{fig:df}
\end{figure}

We should note that this issue was not reported in the Kafka-Direct paper~\cite{kafkadirect}, which evaluated an experiment in which the publisher and subscriber ran on the same node with the subscriber polling for new data.  Kafka-Direct deploys one thread per client node, hence with both on the same client node, the publisher and subscriber are handled by the same thread and take turns.  In our experiment publishers and subscribers ran on {\em different nodes} (we believe this is more realistic), resulting in two threads that contend for a lock, causing the high and variable latencies we observed.  A batched-get feature enables high throughput (albeit lower than for the CMS).  In contrast, with our CMS design subscribers await broker notifications, and lock contention does not arise.

\subsection{A Smart Farming Application}
\label{sec:sf_app}\label{sec:sf_eval}

Next we developed a dairy health tracking application that images animals as they
enter or leave a milking parlor. ML is then employed to develop a variety of
information streams reported via dashboards to farm workers, owner and vets,
resulting in a multi-stage pipeline. The first step uses a motion detector and
RFID to capture photos and identify the cows.  The second step fully analyze
selected images, after which web tools generate dashboard reporting.

Two vision models are involved in this application: a filter model that determines whether a photo has a valid animal image, and a body-condition scoring (BCS) model called CowNet~\cite{cownet} that assesses the health condition of the animal. We developed ${\lambda}_{filter}$ and ${\lambda}_{bcs}$ as DLL lambdas on Cascade, using TensorFlow's C API to incorporate the two models.  A storage stage is represented as a no-op lambda, giving the three-stage DFG shown in Fig.~\ref{fig:sf_dfg}. Both $/sf/detect\_animal$ and $/sf/assess\_mastitis$ pools are volatile, since not all original photos are worth saving.  The $/sf/save\_image$ pool is used to retain photos along with the computed body condition score, RFID, and timestamp. These lambdas use round-robin dispatching because they do not have any constraint on ordering. 
Nodes in the front-end and compute stages perform image analysis and run on servers of type A.
The other nodes run in the type B servers.  Each node runs in a dedicated server to avoid resource conflicts.

We first deployed the application with a configuration where each stage of the pipeline runs on a single-member shard.
The raw image size in JPEG format is about 200 KBytes.
At the beginning of the experiment, the photo aggregator transforms the raw images into two-dimensional dense arrays, which can be used directly by the inference engine. 

We then rebuilt the application as a Flink pipeline.
It consists of a photo streaming data source task as the photo aggregator, four filter tasks, four bcs tasks, and a terminating data sink.
We use Flink’s fine-grained resource management to control the task layout so that tasks of the same type will run on the same server.
Moreover, the filter and bcs tasks are placed in Type A servers because the models require GPU resources.
To mimic Cascade in-memory storage of the results, the Flink data sink task saves output into an in-memory hashmap.
Even though Flink's data storage sink is non-replicated and the Cascade version replicates the output, Flink is substantially slower.

\textbf{Latency Breakdown:} In this experiment we selected fixed sending rates
ranging from 50 to 400 frames per second (fps).
For each rate, we run a session that lasts for (at least) five seconds and log the timestamps for each photo at different stages in the pipeline.
We recorded the following six timestamps for each photo:  (1) the photo aggregator sends a photo; (2) the filter lambda is triggered; (3) the filter lambda terminates; (4) the bcs model is triggered; (4) the bcs model terminates;  (5) the result is written by the store pool.

The gray bars in Fig.~\ref{fig:dflat} show results for Cascade at a low rate and at the highest sustainable rate (the limiting factor is teh cost of the filter and bcs models, which max out our type A servers).
The end-to-end latency is only six milliseconds for the light workload at $\sim 50$ fps and the stressed workload at $\sim 400$ fps.
Even when the workload grows to a stressed 650 fps, it only rises slightly to 6.5 milliseconds.
Model inference time dominates end-to-end latencies: 
filter processing time represents about 40\% of end-to-end latency, and bcs model processing is even greater, at nearly 43\%.
The aggregated data forwarding latency is just 17\%, reflecting the efficiency of the data path.

The yellow bars in Fig.~\ref{fig:dflat} show the Flink latency breakdown.
Although the filter and bcs models consume slightly more time, Flink's data forwarding delays (highlighted with stripes) are far higher.
For a stressed load at 400 fps, Cascade's end-to-end latency is about one-eighteenth that of Flink's.
Even with light load at 50 fps, Flink's end-to-end latency is 25 $ms$, whereas Cascade is just 6 $ms$, a 75\% reduction.
The peak achievable Flink throughput was ${\sim} 450$ fps, hence there is no Flink data point for $650$ fps.

\textbf{Throughput Scalability:} We investigated scalability by varying the number of nodes while tracking throughput.
Here, the configuration of the shard responsible for each stage has a significant impact, so we use a tuple to represent a system configuration, where the elements are positive integers representing the number of nodes assigned to each role:
frontend (which runs the filter lambda), compute (which runs the bcs scoring lambda).
Since the storage is not a performance bottleneck here, we keep the same storage tier set-up as in the previous section (three nodes in the storage shard) in this experiment and skip it in the tuple.
For example, $(1,2)$ represents a system configuration with six nodes.  The frontend pool is backed by a shard with one node; the compute pool is backed by a shard with two nodes; and, not shown in the tuple, the store pool is backed by a shard with three nodes.
We then graphed the maximum throughput achievable without overloading the pipeline in Fig. ~\ref{fig:dfthp}.
For context we  benchmarked both lambdas on a single type A server: filter runs at $\sim 900$ fps, while bcs runs at $\sim 700$ fps.

The overall trend for Cascade, shown in dark gray bars in Fig.~\ref{fig:dfthp} is easily understood.
The bcs lambda is the bottleneck in the $(1,1)$ configuration.
In the $(1,2)$ configuration, the bcs lambda has adequate capacity because it runs on 2 nodes, causing the filter lambda to emerge as the limiting factor:
we obtain a maximum throughput of $\sim 900$ fps.
With configuration $(2,2)$  bcs is again the limit.
With the $(2,3)$ configuration, the two lambdas are balanced and throughput exceeds 1000 fps.
Broadly, these results support our view that Cascade has excellent scalability.
We repeat the same experiment using Flink (yellow bars).  
Cascade turns out to outperform Flink by $\sim 40\%$, reflecting the benefits of our architecture.

\subsection{Real-time Collision Detection}
\label{sec:rcd_eval}
\label{sec:rcd_app}
Finally, we created an application to  make a traffic intersection aware of hazards.  For example, it might be equipped with bright red warning lights on the traffic poles, flashing them if a collision is imminent.

We analyze video streams that include several types of agents (pedestrians, bikers, cars, etc).  The solution extrapolates agent trajectories to detect collisions before they occur, using  off-the-shelf ML models  (Fig.~\ref{fig:rcd_dfg}).  The first stage, ${\lambda}_{mot}$, runs a Multi-Object Tracking (MOT) ML model to track trajectories. 
${\lambda}_{YNet}$  predicts each trajectory for the next 4.8 seconds based on the past 3.2 seconds.
${\lambda}_{detect}$ predicts potential collisions and requires 
 consistency: stale data could disrupt the algorithm.  The final stage stores output.   

${\lambda}_{mot}$ uses a multi-object tracker available in~\cite{yolov5-strongsort-osnet-2022}, which combines YOLO5 for agent detection, and StrongSORT~\cite{strongsort} and OSNet~\cite{osnet} for trajectory tracking.
${\lambda}_{YNet}$ use a model called YNet~\cite{ynet}.
Both ${\lambda}_{mot}$ and ${\lambda}_{YNet}$ runs in PyTorch.
About 200 lines of Python code were required to wrap
each ML model in a DLL lambda. The lambda API was used to retrieve the
required input objects through get operations.
${\lambda}_{detect}$ performs a linear interpolation on each trajectory prediction and checks if any interpolated pair of trajectories in the same frame crosses each other: a potential collision.
All ML models used in ${\lambda}_{mot}$ and ${\lambda}_{YNet}$ are trained with the Stanford Drone Dataset (SDD)~\cite{sdd}. 
In our experiment, we treat each video as a stream originating at a camera hosted by a distinct client. 

Next, we evaluated the traffic safety application. Ideally, computations (${\lambda}_{mot}$, ${\lambda}_{YNet}$, and ${\lambda}_{detect}$)
should dominate the end-to-end latency of each video frame.
${\lambda}_{mot}$ was deployed on one shard containing two type A servers. ${\lambda}_{YNet}$ was
deployed on one shard containing six type A servers. Since ${\lambda}_{detect}$ does not require
GPUs, it was deployed on one shard containing three type B servers. Finally, we deployed three
clients on type B servers; each streams one randomly selected video from the SDD
dataset.

Clients stream videos in a rate of 2.5 frames per second. Each frame is sent uncompressed to the first stage of the pipeline, and has a size of
roughly 8MB. 
We executed the application with all three clients streaming simultaneously for approximately 5 minutes. The first 30 seconds worth of frames from each stream were discarded to allow a warmup, leaving an aggregated total of 2073 frames.
${\lambda}_{mot}$ typically detected 7 to 16 agents per frame.

\begin{figure}[t]
  \centering
  \includegraphics[width=\linewidth]{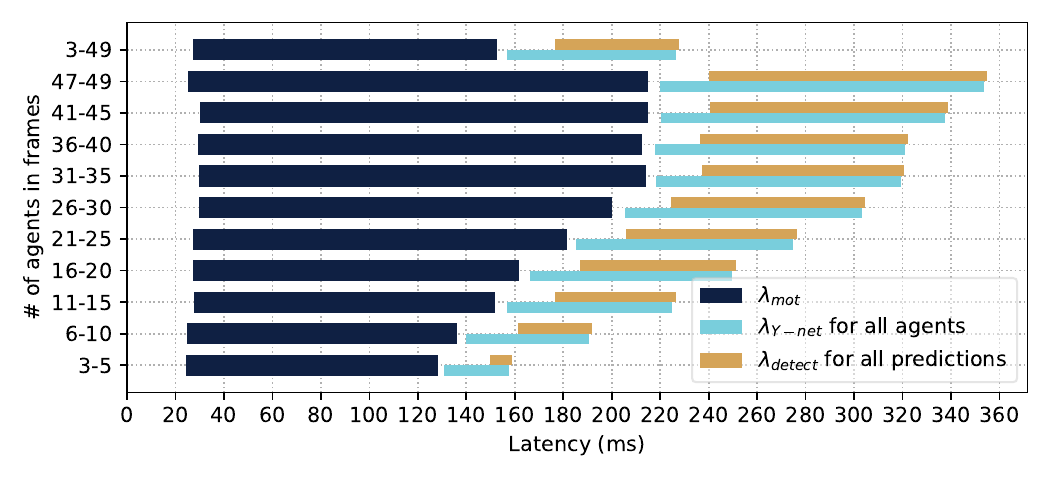}
  \caption{RCD average Per-Frame Latency Breakdown}
  \label{fig:rcd_lat}
\end{figure}

Of particular interest is the end-to-end frame-processing latency and the cost of Cascade lambdas.  
Fig.~\ref{fig:rcd_lat} shows the average latency breakdown per frame. The Y axis aggregates
frames according to the number of agents detected, in groups of five. The topmost group corresponds to all frames. The X axis indicates latency in
milliseconds.  Each horizontal bar corresponds to the average end-to-end latency for a lambda in
frames containing the indicated number of agents. In a frame, the end-to-end latency for
${\lambda}_{mot}$ corresponds to the time since the lambda started executing (with the frame already
available in memory), until all agents in the frame are detected and their trajectories computed.
Since there are multiple agents per frame, the end-to-end latency for ${\lambda}_{YNet}$
corresponds to the time since the first instance of this lambda starts, until the last instance
finishes. Multiple instances are executed, generating predictions in parallel. Thus
multiple instances of ${\lambda}_{detect}$ start while there are still instances of
${\lambda}_{YNet}$ running. The end-to-end latency for ${\lambda}_{detect}$ is the time since the
first instance started, until the last instance finishes. The gap between time 0ms and
${\lambda}_{mot}$ corresponds to the frame transfer from the client. The gap between
${\lambda}_{mot}$ and ${\lambda}_{YNet}$ includes the transferring of the new agents' positions, as
well as waiting time due to the servers being busy executing ${\lambda}_{YNet}$ for the previous
frame.

As expected, latency is higher in general for larger workloads (indicated by the number of agents).
However, the time between when the frame is sent by a client until
${\lambda}_{mot}$ starts is consistent regardless of the workload: the average is 28ms, with a 95th
percentile of 34ms. Similarly, the overhead between ${\lambda}_{mot}$ and the first instance of
${\lambda}_{YNet}$ is consistent, about 5ms.   Focusing on the frames containing 11 to 15 agents, the average end-to-end latency for
the pipeline is 229ms, with a 95th percentile of 264ms.  Considering all frames, the latency
is much less consistent with an average of 241ms and a 95th percentile of 338ms. This 
variance stems from lambdas that require more processing time when the workload is
heavier.  We conclude that Cascade incurs a low and consistent overhead in the critical
path.

We then deployed the identical logic on Microsoft Azure Cloud.  Adhering closely to
documentation~\cite{asaandml}, we created a pipeline using Azure Machine
Learning (AML), Stream Analytics (SA), and Event Hubs (EHs). Application
lambdas were deployed as real-time endpoints (which behave as web-services),
triggered by SA jobs connected by EHs~\cite{kettner2022iot}.   The deployment effort was 
comparable: about 200 lines of code, consisting of Python scripts and configuration
files, per ML model.

In this experiment, we employed only one camera, and preloaded the videos to Azure Blob Storage~\cite{calder2011windows}.
The simulated camera runs in a virtual machine, triggering the application pipeline by sending
frame metadata to an EH at a rate of 2.5 fps. A front-end SA job invokes ${\lambda}_{mot}$, which downloads the frame from the Blob storage and
initiates object tracking. Results are then passed through another EH. The third job
was triggered similarly. The results of the last job are returned to the camera virtual
machine for end-to-end latency measurement.

We used seven $\textit{Standard\_NC4as\_T4\_v3}$ AML instances equipped with the same
GPU as in our local environment, one for ${\lambda}_{mot}$ and six for ${\lambda}_{YNet}$.
Three more instances (type $\textit{Standard\_DS3\_v2}$) were deployed for ${\lambda}_{detect}$.
We employed premium-tier~\cite{azehns} EHs, and SA jobs ran in a dedicated SA cluster. Measurements confirmed that our lambdas have the same compute costs as with Cascade.

\begin{figure}
    \centering
    \subfloat[End-to-End\label{fig:azureapp}]{
        \includegraphics[width=0.53\linewidth,,]{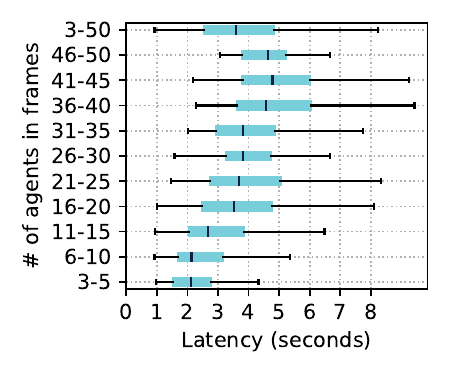}
    }
    \subfloat[Event Hub\label{fig:azehlat}]{
        \includegraphics[width=0.44\linewidth,,]{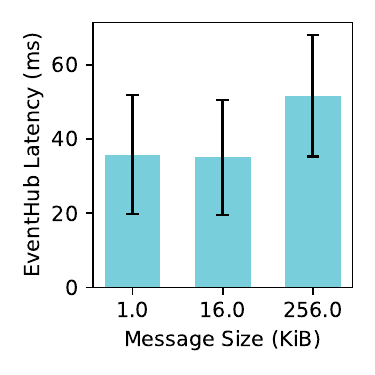}
    }
    \caption{Latency on Azure Cloud}
    \label{fig:azperf}
\end{figure}

Although Azure solutions are capable of high throughput, it is quite difficult to disable batching in standard Azure components. Figure~\ref{fig:azureapp} shows end-to-end
latency, grouped by the number of agents detected in a frame. The topmost bar corresponds to all
frames. Medians are all above 2 seconds, with long tails. The minimum was 924.67ms. 
Figure~\ref{fig:azehlat} drills down, showing that the latency originates in the EH: We deployed a consumer and a publisher in an idle Azure virtual machine to stress a premium-tier
EH with a single partition. Latencies from sending to receiving is tracked for 30 seconds, using the same message sizes as in Figure~\ref{fig:kafka-direct}. We 
disabled batching and controlled the sending rate for the lowest latency. The average EH latency is
tens of milliseconds, over two orders higher than RDMA-enabled CMS, and jitter is common (note the long tails in Figure~\ref{fig:azureapp}).   

\section{Additional Related Work}
Sections 1 and 2 discussed  a number of widely used big-data platforms and the challenges of adapting them for use in event-driven edge settings.  Although Kafka Direct, Apache Flink and Storm~\cite{flink} aim at stream processing, these are not the only prior systems relevant to our effort.  Spark~\cite{spark} achieves impressive performance for iterative tasks such as training.  It gains this speed through in-memory RDD caching and scheduling, but runs the actual jobs on nodes distinct from the HDFS storage service that hosts data.  Cascade is similar in style, but with a stronger bias towards low-latency that led us to group related objects and run computations close to their inputs.

Prior work on K/V stores includes RDMA-enabled systems such as FARM~\cite{FaRM} and FASST~\cite{FaSST} as well as commercial data warehousing products, such as Amazon's DynamoDB~\cite{DynamoDB}, Snowflake ~\cite{snowflake}, Microsoft CosmosDB, Databricks Datalake, Cassandra, RocksDB or even the Ceph object-oriented file system, which runs over a key-value store called RADOS~\cite{Ceph}.  As with Spark, none of these solutions hosts developer-supplied lambdas or deploys GPU accelerators close to the storage system, forcing costly locking, copying and domain crossings.  Moreover, few focus on rapid data consistency (Azure has long featured strong storage consistency, and AWS recently introduced a consistency feature, but neither achieves particularly low update latency).   Cascade lacks support for complex transactions but does provide a form of linearizability: each get or put is treated as a separate event.  Puts run atomically, and get only sees stable (committed) data.  

Closer to our approach is RedisAI, a system that combines a K/V store with ML model serving, outperforming TensorFlow and PyTorch\cite{9652868,pieterannouncing}. However, RedisAI lacks RDMA support and other data-path optimizations, such as the Cascade trigger put.   Moreover, a RedisAI application DAG is handled by one node whereas a Cascade DFG can scale over multiple nodes.
\section{Conclusion}
We created Cascade to host a new generation of edge computing that depends on large stored objects and other ``big data'' collections, yet also must carry out computations under time pressure.  AI developers depend on modern cloud tools and platforms, hence any new option must be as transparent as possible.  Cascade's architecture achieves these goals, allowing user code to execute on the same machines that run our servers and positioning computation side-by-side with any required data.  The design enables end-to-end zero copy RDMA data paths between the application and our storage model, or between application stages.  Performance is excellent: stage to stage delays can be as low as 33$\mu s$ and bandwidth as high as 4.5GBps.    The latency figure improves on today's standard platforms by multiple orders of magnitude, while the throughput equals or improves upon what today's platforms achieve.

Although Cascade's K/V data model is simple, such stores can still hold complex objects and tables.  Contextualized MLs are likely to leverage this store, issuing embedded query operations under time pressure.  Cascade K/V performance is excellent, and the system guarantees linearizability: an important property in reactive edge settings where inconsistency can result in visible errors or real-world harm.  

\section{Acknowledgments}
We are grateful to Professor Julio Giordano, Martin Matias Perez and Daniel Martin for their help in designing the intelligent dairy pipeline we described here. Yueing Li, Chris de Sa, and Bharath Hariharan made many valuable suggestions early in our design process, and Ranveer Chandra suggested some use cases that shaped the overall solution.  Partial support for our work was provided by AFRL/RY under the SWEC program, Microsoft, Nvidia and Siemens.

\bibliographystyle{ACM-Reference-Format}
\bibliography{cascade}

\end{document}